\documentclass[superscriptaddress,prl,twocolumn,amsmath,amssymb]{revtex4-1}
\usepackage{graphicx}
\usepackage{dcolumn}
\usepackage{bm}
\usepackage{eurosym}
\usepackage{amsmath}
\usepackage{amssymb}
\usepackage{dcolumn}
\usepackage{bm}
\graphicspath{{../images/}}
\usepackage{soul}
\usepackage{braket}
\def\>{\rangle}
\def\<{\langle}
\usepackage[bottom]{footmisc}

\begin{document}

\title{High-dimensional entanglement certification} 

\author{Zixin Huang}
\affiliation{Quantum Photonics Laboratory, School of Electrical and Computer Engineering, RMIT University, Melbourne, Australia and School of Physics, University of Sydney, NSW 2006, Australia}

\author{Lorenzo Maccone}
\affiliation{Dip. Fisica and INFN Sez. Pavia, University of Pavia, via Bassi 6, I-27100 Pavia, Italy}

\author{Akib Karim }
\affiliation{Quantum Photonics Laboratory, School of Electrical and Computer Engineering, RMIT University, Melbourne, Australia and School of Physics, University of Sydney, NSW 2006, Australia}

\author{Chiara Macchiavello}
\affiliation{Dip. Fisica and INFN Sez. Pavia, University of Pavia, via Bassi 6, I-27100 Pavia, Italy}

\author{Robert J. Chapman}
\affiliation{Quantum Photonics Laboratory, School of Electrical and Computer Engineering, RMIT University, Melbourne, Australia and School of Physics, University of Sydney, NSW 2006, Australia}

\author{Alberto Peruzzo}
\affiliation{Quantum Photonics Laboratory, School of Electrical and Computer Engineering, RMIT University, Melbourne, Australia and School of Physics, University of Sydney, NSW 2006, Australia}
\begin{abstract}
  Quantum entanglement is the ability of joint quantum systems to possess global properties (correlation among systems) even when subsystems have no definite individual property. Whilst the 2-dimensional (qubit) case is well-understood, currently, tools to characterise entanglement in high dimensions are limited.  We experimentally demonstrate a new procedure for entanglement certification that is suitable for large systems, based entirely on information-theoretics. It scales more efficiently than Bell's inequality, and entanglement witness.  
  The method we developed works for arbitrarily large system dimension $d$ and employs only two local measurements of complementary properties. This procedure can also certify whether the system is maximally entangled.   We illustrate the protocol for families of  bipartite states of qudits with dimension up to $32$ composed of 
  polarisation-entangled photon pairs. 
\end{abstract}

\maketitle
%
%
\thispagestyle{empty}

\section*{Introduction}
As the dimension of investigated systems increases, it becomes more
complicated to demonstrate their quantum effects \cite{arndt2014testing, sciarrino2013quantum, friedman2000quantum}. Indeed, a full characterization (quantum tomography) becomes practically impossible already for systems of rather small dimensionality \cite{PhysRevA.66.062305, artiles2005invitation}. It is therefore important to explore new avenues to prove the presence of such effects, e.g.~entanglement, for arbitrary dimensions.  High-dimensional entangled states offer a larger code space, attracting interests for quantum key distribution \cite{mair2001entanglement}, teleportation \cite{PhysRevLett.69.2881} and security-enhanced quantum cryptography \cite{PhysRevLett.88.127901, PhysRevA.69.032313}.

 In this context, proving that one has achieved entanglement (entanglement certification) and
detecting entanglement are different primitives.  Indeed, entanglement
detection methods \cite{bruss2002characterizing, RevModPhys.81.865, Ghne20091} must be as sensitive as possible and must be able to detect the largest possible class of entangled states. Often such methods are inapplicable to large system dimensions or scale poorly with increasing dimension as they entail increasingly
complicated measurements and data analysis. Entanglement detection  such as witness operators, for instance, typically requires a number of local measurement settings that scales linearly in $d$ \cite{guhne2003experimental}.
 In contrast, entanglement
certification refers to the fact that one has simply to prove that the system is
entangled. To do entanglement certification, one can optimize the
method to the specific entangled state that one is producing. It must fulfill different requirements: it must be as
robust and simple as possible.   In other
words a good method for entanglement detection can also work for
entanglement certification, but not vice versa.

\begin{figure}[h!]
\includegraphics[trim = 0cm 0cm 0cm 0cm, clip, width=8cm]{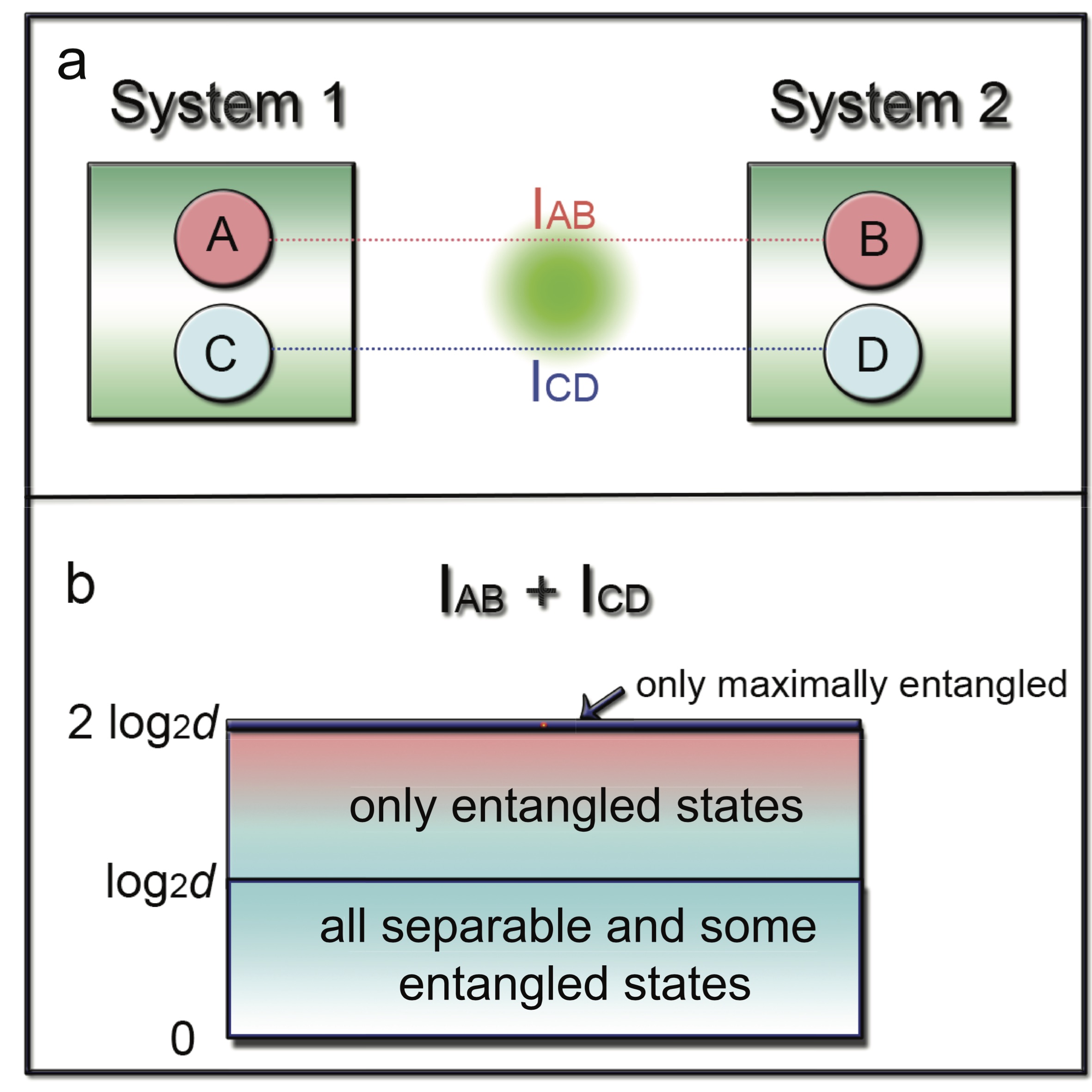}
\caption{\label{f:fig1} Illustration of the procedure.  (a)~Measure two complementary
  observables: $AB$ and $CD$, where $A, C$ are observables for system
  1 and $C, D$ for system 2. Then calculate the mutual information
  $I_{AB}$ and $I_{CD}$ of their outcomes. (b)~If the value of the sum
  $I_{AB}+I_{CD}$ is larger than log$_2(d)$ bits ($d$ being the system
  dimension), then the two systems are certified to be entangled. If
  the value of the sum is 2log$_2(d)$ bits, they are certified to be
  maximally entangled.}
\end{figure}

Here we present an entanglement certification procedure that is extremely simple to implement (the
measurement of two local observables \cite{PhysRevA.75.012336, PhysRevA.75.062317} suffices), is compatible with current state-of-the-art experimental techniques, and can be easily
scaled up to arbitrary dimension. In addition to certifying the
production of entangled states, our procedure can also certify the
production of maximal entanglement. To prove its simplicity, we
present an experimental test that uses entangled systems with
dimension up to $d=32$, constructed by suitably grouping couples of
entangled photon pairs. In the general case, for a two-qudit experiment with arbitrary $d$,
two measures would still be sufficient to implement our method, but not sufficient for 
tomography or other entanglement detection methods. To certify the presence of entanglement, one
has only to calculate the (classical) correlations among the
measurement outcomes of the two observables, for example, through
their mutual information. If such correlations are larger than some
threshold, the state is guaranteed to be entangled
(Fig.~\ref{f:fig1}).  If they attain their maximum value, the state
must be maximally entangled.

\begin{figure*}[hbt]
\includegraphics[ width=1.0\linewidth]{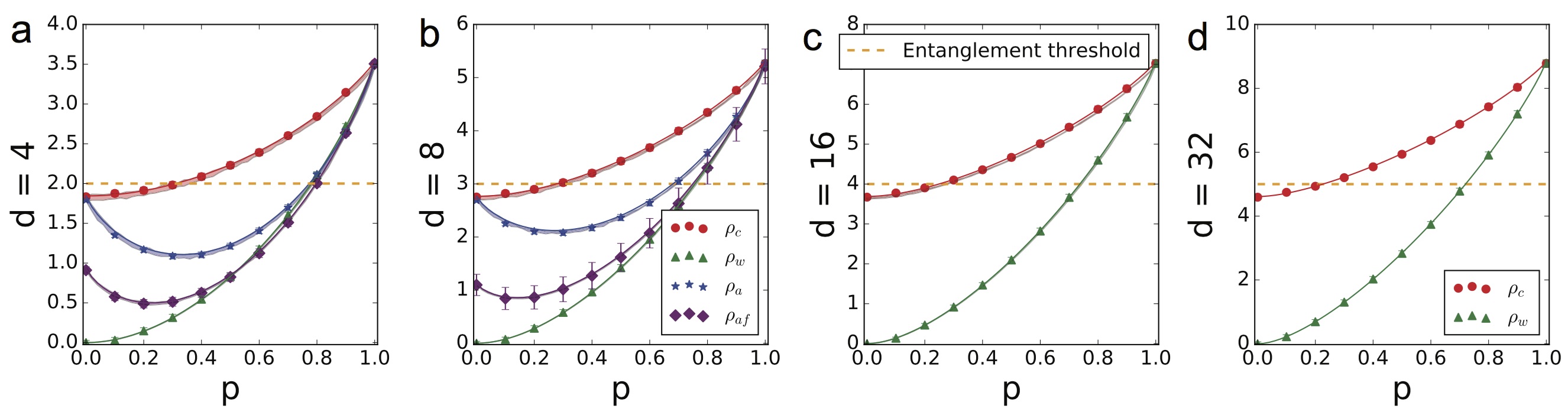}
\caption{\label{fig:higher_d} Experimental test of the method for
  $d$-dimensional systems. The plots refer to the experimentally
  determined sum of mutual information $I_{AB}+I_{CD}$ for $\rho_c$ of
  equation \eqref{st1} (red circles), $\rho_w$ of equation \eqref{eq:werner} (green
  triangles), $\rho_a$ of equation \eqref{eq:rho6} (blue stars when $CD$ is
  the $\sigma_x$ observable, and purple diamonds when $CD$ is the
  Fourier basis). The points that fall above the entanglement
  threshold (orange dashed line) refer to states that are
  \emph{certified} to be entangled. Taking into account the confidence interval of the measurement outcome, the state was maximally entangled if the  maximum sum $I_{AB}+I_{CD}=2\log_2d$ is achieved. The bands are a Monte Carlo simulation, taking into
  account realistic experimental conditions (eg. phase inaccuracy due
  to the wave plates used in state tuning); for $d = 32$ they are narrower than the fit. The error bars represent bounds for 2 standard deviations
  (95$\%$ confidence), for most of the data points the error bars are
  smaller than the markers. }
\end{figure*}

\begin{figure}[h!]
\includegraphics[ width=8cm]{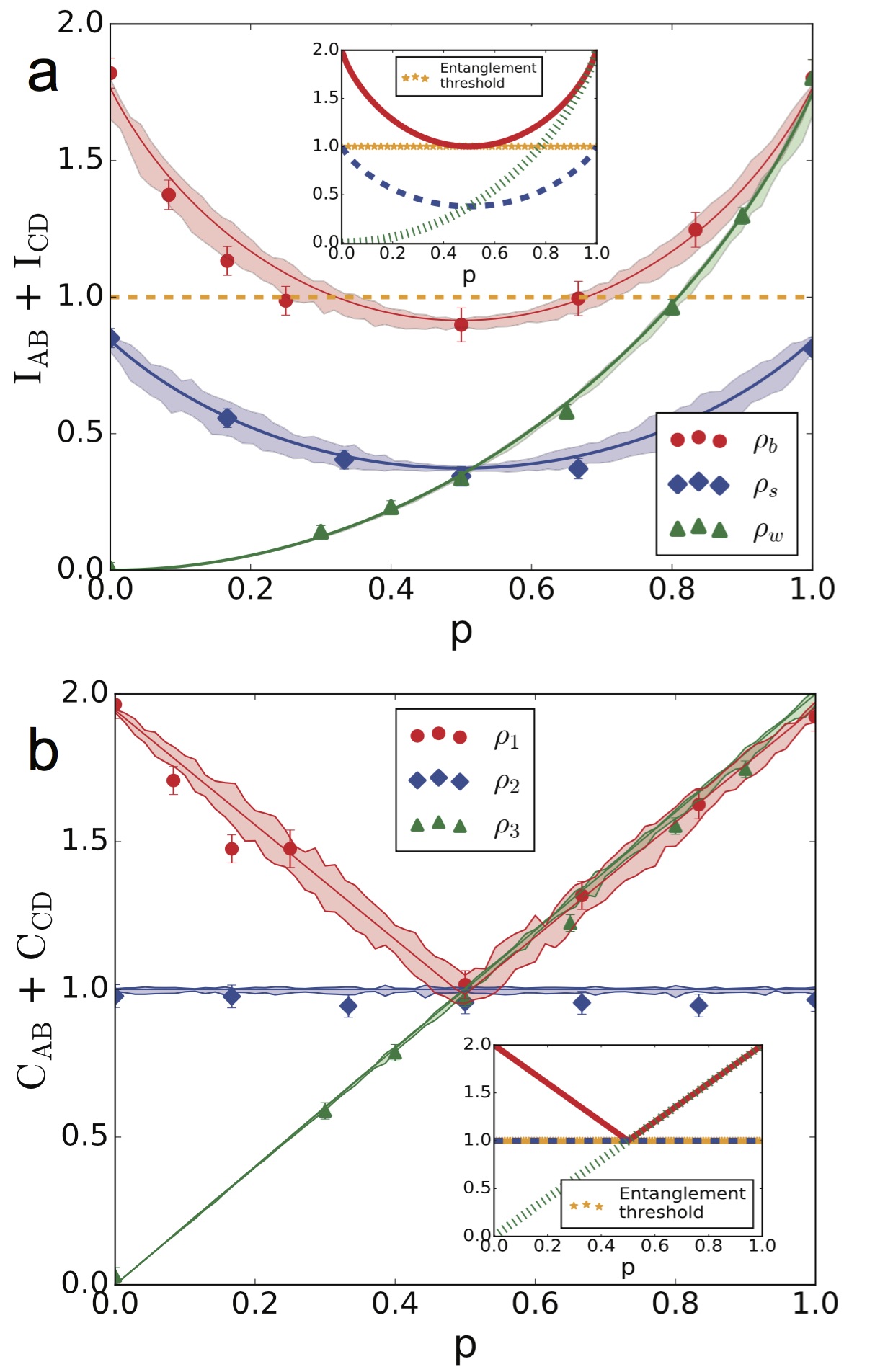} 
\caption{\label{fig:pearson} Experimental comparison between mutual
  information (above) and Pearson coefficient (below). The insets show
  the theoretical curves. The data points refer to $\rho_b$ of equation 
  \eqref{eq:bell} (red circles), $ \rho_s $ of equation \eqref{eq:classical}
  (blue diamonds), $ \rho_w $ of equation \eqref{eq:werner} (green triangles) for $d = 2$. The
  states whose data points are strictly above the entanglement
  threshold ($I_{AB}+I_{CD}=1$ for the upper graph and
  $C_{AB}+C_{CD}=1$ for the lower graph, conjectured) are certified to be entangled. In
  the lower graph this threshold overlaps with the fit for $\rho_s$,
  which is then not entangled, as expected. The markers represent the
  data points, the lines represent a fit from reconstructed density
  matrices, and the bands are a Monte Carlo simulation, taking into
  account realistic experimental conditions (eg. phase inaccuracy due
  to the wave plates used in state tuning).  The error bars represent
  bounds for 2 standard deviations. It is evident that the
  experimental results are in good agreement with the theoretical
  values.  }
\end{figure}

If $A$ and $C$ are complementary properties, the knowledge of $A$
gives no information on $C$ and vice versa. This happens whenever
$|\braket{a|c}|^2=1/d$, for all $|a\>$ and $|c\>$ eigenstates of the
observables $A$ and $C$, where $d$ is the Hilbert space dimension.
This is equivalent to two sets of mutually unbiased bases (MUB's).
Consider the two-qubit maximally 
entangled state $|\Phi^+\>\propto|00\>+|11\>$; it clearly has maximal
correlation among results for the observables with eigenstates
$|0\>,|1\>$: both qubits have the same value. This state has also maximal
correlation among results for a complementary observable with
eigenvalues $|\pm\> = \frac{1}{\sqrt{2}}(|0\>\pm|1\>)$, since it can be written as
$|\Phi^+\>\propto|++\>+|--\>$. Thus, the mutual information between
measurement outcomes on the two qubits is one bit per observable,
summing to two bits. If one has only separable states, the sum
cannot be larger than one: for example, the classically correlated
state $\rho \propto \ket{00}\bra{00}+ \ket{11}\bra{11}$ has one bit of mutual
information for the outcomes of the first observable, but zero bits
for the outcomes of the second. 

 Starting from the theoretical suggestions of \cite{maccone2014complementarity}, we extend the method from the qubit ($d=2$) case to arbitrary high dimensions. The proposed mechanism to certify entanglement thus uses the following
procedure:
\begin{enumerate}
 \item Identify two bipartite complementary
  observables $AB$ and $CD$, where $A,C$ are system 1 observables and
  $B,D$ are system 2 observables (Fig.~\ref{f:fig1}a).
 \item Measure
  the statistics of the outcomes of the two observables: local
  measurements on the two systems suffice.
 \item These
  measurements return the joint probabilities $p(a_o, b_o)$ of
  obtaining outcome $ a_o$ on system 1 for $A$ and $ b_o$ on
  system 2 for $B$, and $p( c_o, d_o)$ of obtaining outcome $
  c_o$ on system 1 and $ d_o$ on system 2. Use them to calculate the
  mutual information $I_{AB}$ among measurement results for $AB$ and
  $I_{CD}$ among results for $CD$, with
  \begin{eqnarray}
  I_{AB}\equiv\sum_{ a b}p( a, b)\log_2\frac
  {p( a, b)}{\sum_{a}p( a, b)\sum_{b}p( a, b)}
  \label{mi}.
  \end{eqnarray}

\item Certification: if $I_{AB}+I_{CD}>\log_2d$, the two systems are
  entangled; if $I_{AB}+I_{CD}=2\log_2d$, the two systems are
  maximally entangled.
\end{enumerate}

The proof of these statements, based on Maassen and Uffink's entropic uncertainty relation \cite{PhysRevLett.60.1103} is given in \cite{maccone2014complementarity} (see also \cite{PhysRevA.90.062119,PhysRevA.89.022112, PhysRevLett.108.210405}).

In regards to the choosing the observables, for pure states, the obvious choice would be the Schmidt bases and their respective Fourier bases. 
Whilst for mixed states, one can diagonalise the density matrix, identify the eigenvector
with the largest weight and use the Schmidt basis (and its respective Fourier bases). Alternatively, one can choose the bases that diagonalises the reduced density matrices. While there is no guarantee that these choices will allow one to implement the procedure, they are the ones that may uncover the most correlations.

This method is simple to implement for systems of any dimensionality:
it only entails independent measurements of two local observables on the two
systems. Moreover, it is robust since, although one can optimize the
choice of observables to maximize the sum of mutual information, the
systems are guaranteed to be entangled if the above conditions are
satisfied for {\em any} couple of complementary observables. 
It is interesting to note that, coherently performing sequential 
complementary measurements on the same system may generate the entanglement itself \cite{PhysRevA.89.010302}.

To date, the most prominent way of producing higher dimensional entangled systems is via the orbital angular momentum degree of freedom of a photon \cite{fickler2012quantum,dada2011experimental,mair2001entanglement} ; schemes to produce three-level entangled states in trapped ions have also been proposed \cite{PhysRevA.68.035801,PhysRevA.77.014303}. Our method, however, works for any d-dimensional system as long as the appropriate measurements can be performed.

Based on the theoretical work by Collins et al., other recent experiments \cite{dada2011experimental,lo2015experimental} have studied higher dimensional entangled system via generalised Bell's inequalities, where the correlations between the two measurement settings on each are studied. The violation of Bell's inequality is usually discussed in terms of quantum non-locality. In this context, the framework of hidden variable theories is in general specified by the 4 measurement settings, each with $d$ outcomes, therefore needing $d^4$ joint probabilites \cite{collins2002bell} to describe the system globally. In this case,  \cite{collins2002bell,dada2011experimental,lo2015experimental} examine two settings on each system, and how the observables on one system are correlated with the two observables on the second system, requiring a total of $2d\times 2d $ joint outcomes. On the other hand, our method is based on entropic relations, requiring only two measurement settings, and is completely specified by $2d^2$ probabilities. Similarly to \cite{lo2015experimental}, we used ensembles of individual entangled
photon pairs to construct a higher dimensional state. All the subsystems that compose this state do not exist at the same time, but
that is irrelevant to our current aims.

To demonstrate the method in practice, we performed an experiment
using high-dimensional entangled systems. We certify the presence of
entanglement for various families of states, by measuring pairs of
complementary observables. These families are obtained by
appropriately grouping couples of polarisation entangled photonic qubits, mapping qudits onto qubits:
\begin{eqnarray}
\frac1{\sqrt{d}}\sum_{j=0}^{d-1}\ket{j} \ket{j}&=&\frac1{\sqrt{d}}\Big(\ket{00}+\ket{11}
\Big)^{\otimes
  n},\label{mappings1}
\end{eqnarray}
where this mapping is obtained by expressing $j$ in binary notation
and rearranging the qubits in such a way that the ones placed in odd
positions in the tensor products on the right hand side are assigned
to system 1 while the ones in even positions to system 2. 
For example, a $d=4$ maximally entangled state can be expressed
as $(\ket{00} + \ket{11} + \ket{22} +\ket{33})/2 \equiv (\ket{0 0\,00}
+ \ket{01\, 01} + \ket{10\, 10} + \ket{11\, 11})/2\equiv
(\ket{0}_\alpha\ket{0}_{\beta} + \ket{1}_\alpha\ket{1}_{\beta})
\otimes (\ket{0}_\gamma\ket{0}_{\delta} +
\ket{1}_\gamma\ket{1}_{\delta})/2,$ where qubits labeled $\alpha$ and
$\gamma$ belong to system 1, while $\beta$ and $\delta$ to system 2.
It would also be extremely interesting to show that one could coherently manipulate the $n$ entangled pairs locally, though that is not necessary to implement our method (which is also one of the main strength of the method). An analogous mapping applies to the mixed states:
\begin{eqnarray}
  \tfrac1{{d}}\textstyle{\sum_{j}}\ket{j}\bra{j}\otimes\ket{j}\bra{j}
  &=&\tfrac1{{d}}(\ket{00}\bra{00}+\ket{11}\bra{11})^{\otimes
    n}
\label{mapping}.
\end{eqnarray}

Regarding the necessary measurements, the observable $AB$ corresponds
to $\sigma_z$, namely the computational basis
$\{\ket{0},\ket{1}\}$ for each qubit. As to the observable $CD$, we will
consider two possibilities: the Fourier basis and the $\sigma_x$
basis. The Fourier basis is defined as
\begin{eqnarray}
\ket{f_j} \equiv \frac 1{\sqrt{d}}\sum_{k=0}^{d-1}\omega^{kj}\ket{k} 
\Rightarrow
\ket{k}=\frac 1{\sqrt{d}}\sum_{j=0}^{d-1}\omega^{-kj}\ket{f_j}\
\label{fourier},
\end{eqnarray}
with $\omega\equiv\exp(2\pi i/d)$. It can be expressed as tensor
products of single-qubit states (see Supplementary Material).

The Fourier basis in arbitrary dimension identifies an observable
complementary to the computational basis. However, in the case considered here, there are complementary bases that are simpler to access
experimentally, namely the bases where one measures $\sigma_x$ or $\sigma_y$ on each qubit.  We consider $\sigma_x$ here.
 The $\sigma_x$ basis is given by $|c_k\>$ ($k=0,\cdots,d-1$) obtained by expressing the binary digits of $k$ in the
$|+\>,|-\>$ basis, i.e.
\begin{eqnarray}
|c_k\>=[(|0\>+(-1)^{\gamma_1}|1\>)(|0\>+(-1)^{\gamma_2}|1\>)\cdots]/\sqrt{2^n}
\label{sigmaxb},
\end{eqnarray}

\noindent where $\gamma_\ell$ are the bits of the number $k$.  The $\sigma_x$
basis is complementary to the computational basis, since
$|\braket{j|c_k}|^2=1/2^n=1/d$ for all $j,k$.

We tested our entanglement certification procedure on several families
of states. These particular states that we consider are arbitrarily chosen for their simplicity. For the specific examples, the entanglement of the $d$-dimensional entangled pair is exactly the same as the one obtained from consecutive entangled two-photon states. This happens only when $d$ is a power of two, though the method works for any $d$ .  The first is
\begin{eqnarray}
  \rho_c(p)=\frac pd\sum_{jj'}\ket{jj}\bra{j'j'}+\frac{1-p}d\sum_j\ket{jj}\bra{jj}
\label{st1},
\end{eqnarray}
which mixes the maximally entangled state of equation~\eqref{mappings1}
with the state in equation \eqref{mapping} with classical correlation only on the
computational basis, the above state is always entangled for $p\neq 0$. 

The experimental test of this prediction is presented in
Fig.~\ref{fig:higher_d}, red circles (the markers are the data points
and the connecting lines are the expected curves fitted to density
matrices from quantum state tomography). 

The second family of states we consider are the Werner states: a
mixture of a maximally entangled and a maximally mixed state,
\begin{eqnarray}
\rho_w=\frac pd\sum_{jj'}\ket{jj}\bra{j'j'}+\frac{1-p}{d^2}I
\label{eq:werner},
\end{eqnarray}
where $I$ is the identity $d^2 \times d^2$ matrix. These states are
entangled for $p>1/(d+1)$ (e.g.~\cite{PhysRevA.66.062310}).
Their mutual information is experimentally determined as the green triangles in
Fig.~\ref{fig:higher_d}. 

The third family of states is
\begin{eqnarray}
\rho_a\! =\! \frac{p}d\sum_{jj'}\ket{jj}\bra{j'j'}+ (1-p) (\tfrac{\ket{+-}\bra{+-} 
+ \ket{-+}\bra{-+}}{2})^{\otimes n},  \nonumber \\ 
\label{eq:rho6}
\end{eqnarray}
where $\ket{\pm} = \frac{1}{\sqrt{2}}(\ket{0} \pm \ket{1})$.
The experimental results are plotted in Fig.~\ref{fig:higher_d} as
blue stars (when the $CD$ observable is the $\sigma_x$ basis $|c_k\>$)
and as purple diamonds (when $CD$ is the Fourier basis
$|f_j\>$). Measurements in the Fourier basis and in the $\sigma_x$ basis are performed for $d = 4$ and 8 only, due to the
large number of projections required for higher dimensions
 where temporal phase instability would significantly affect the
measurements. This state shows the difference
between the complementary bases: its form implies that correlations
are greater in the $\sigma_x$ than the Fourier basis.

The entanglement certification method can be modified by using a
different measure of correlations. Indeed, instead of the mutual information, one can
also measure correlations with the Pearson coefficient
\begin{equation}
\mathrm{C_{AB} \equiv  ({ \langle AB\rangle - 
\langle A\rangle \langle B\rangle  })/{ \sigma_A \sigma_B}},
\label{eq:pearson}
\end{equation}
\noindent where $\langle X \rangle$ is the expectation value of $X$
and $\sigma_X^2$ its variance.It measures the linear correlation
between the outcomes of two observables $A$ and $B$, and takes values
$-1 \leqslant \mathrm{C_{AB}} \leqslant 1 $. It is conjectured
\cite{maccone2014complementarity} that if $\mathrm{C_{AB} + C_{CD}}>1$, then the state is
entangled.  Moreover, it is known that if $\mathrm{C_{AB} +
  C_{CD}}=2$, the state is maximally entangled.  This may be helpful in high-noise
scenarios, since it allows for the certification of a larger class of
entangled states.  The Pearson
correlation coefficient seems to be more robust against imbalanced
probabilities or decoherence in the experiment (see
Fig.~\ref{fig:pearson}), and numerical simulations suggest that it is
more effective in detecting entanglement \cite{maccone2014complementarity}. So, in addition
to the sum of mutual information $I_{AB}+I_{CD}$, we can use
$\mathrm{C_{AB} + C_{CD}}$ as an alternative way of certifying
entanglement (modulo a conjecture). 

We consider  the two-qubit states
\begin{eqnarray}
\rho_b &=&  p \ket{\Phi^+}\bra{\Phi^+} + (1-p)\ket{\Phi^-}
\bra{\Phi^-}, 
\label{eq:bell}\\
\rho_s&=& p (\tfrac{\ket{00}\bra{00} + \ket{11} \bra{11}}{2}) + 
(1-p)( \tfrac{\ket{++}\bra{++}  + \ket{--}\bra{--}}{2} ),\nonumber\\
\label{eq:classical}
\end{eqnarray}
and the Werner state $\rho_w$ of Eq. \eqref{eq:werner} with $d=2$. In
\eqref{eq:bell}, $|\Phi^\pm\>\equiv({|00\>\pm|11\>})/{\sqrt{2}}$.  The
state $\rho_b$ is entangled for $p \neq 1/2$, whereas $\rho_s$ is
always separable and has zero discord only for $p = 0,1$. For
two-qubit states, the $\sigma_x$ basis $|c_k\>$ coincides with the
Fourier basis $|f_j\>$, so the two possible observables $CD$ we used
above coincide here.

However, we preferred working with mutual information since the fact that the Pearson coefficient can be used to certify entanglement is still a conjecture, whereas it has been rigorously proved for the sum of mutual information.

Whilst it is true that the standard entanglement witness
\begin{equation}
W= I - \sigma_z \otimes \sigma_z - \sigma_x \otimes \sigma_x
\end{equation}
out-performs our method for entanglement dection for the ($d = 2$ case) Werner state (detects entanglement at $p \geq 0.5$), this is not true for arbitrary dimensions in general. 

Finally, also comparing this agains Bell's inequalities using the Werner state, our method using mutual information with two complementary observables (computational and $\sigma_x$ basis), for d = 2, the threshold of log$_2(d)$ is surpassed at p$\approx$ 0.78. However, if we increase this to three observables (indeed one can, and in this case we consider the $\sigma_y$ basis), the entanglement is detected at p$\approx$ 0.65 (see supplementary material of \cite{maccone2014complementarity}), which outperforms Bell's inequalities at $1/\sqrt{2}$ \cite{RevModPhys.81.865}. Also, in the limit of large $d$, our method (see Eq (S7) in the Supplementary Material) also out-performs the threshold in Bell's inequality violation, given in \cite{collins2002bell}.

In summary, we have presented an entanglement certification method that is suitable for large dimensional systems. To our
knowledge, there are no such methods that are as efficient to implement
on a 32-dimensional system such as the one we experimentally study
here.  We have shown that entangled states are more correlated on the
outcomes of complementary observables. One may think that this
property is shared by other types of quantum correlations such as
quantum discord \cite{PhysRevLett.88.017901} but, surprisingly, this is not the case, at least when
the correlations are gauged through  mutual information. Indeed,
the separable state $\rho_s$ has highest $I_{AB}+I_{CD}$ for $p=0,1$
(Fig.~\ref{fig:pearson}), the values for which its discord is null. In
contrast, $I_{AB}+I_{CD}$ is lowest for $p=1/2$ which is where such
state has highest discord.  This unexpected property is lost when the
correlation is gauged with the Pearson coefficient, as
$\mathrm{C_{AB} +  C_{CD}}=1$ for all values of $p$.

\section*{Methods}

\begin{figure}[hbt]
\includegraphics[trim=0cm 1cm 0cm 2cm,clip, width=10cm]{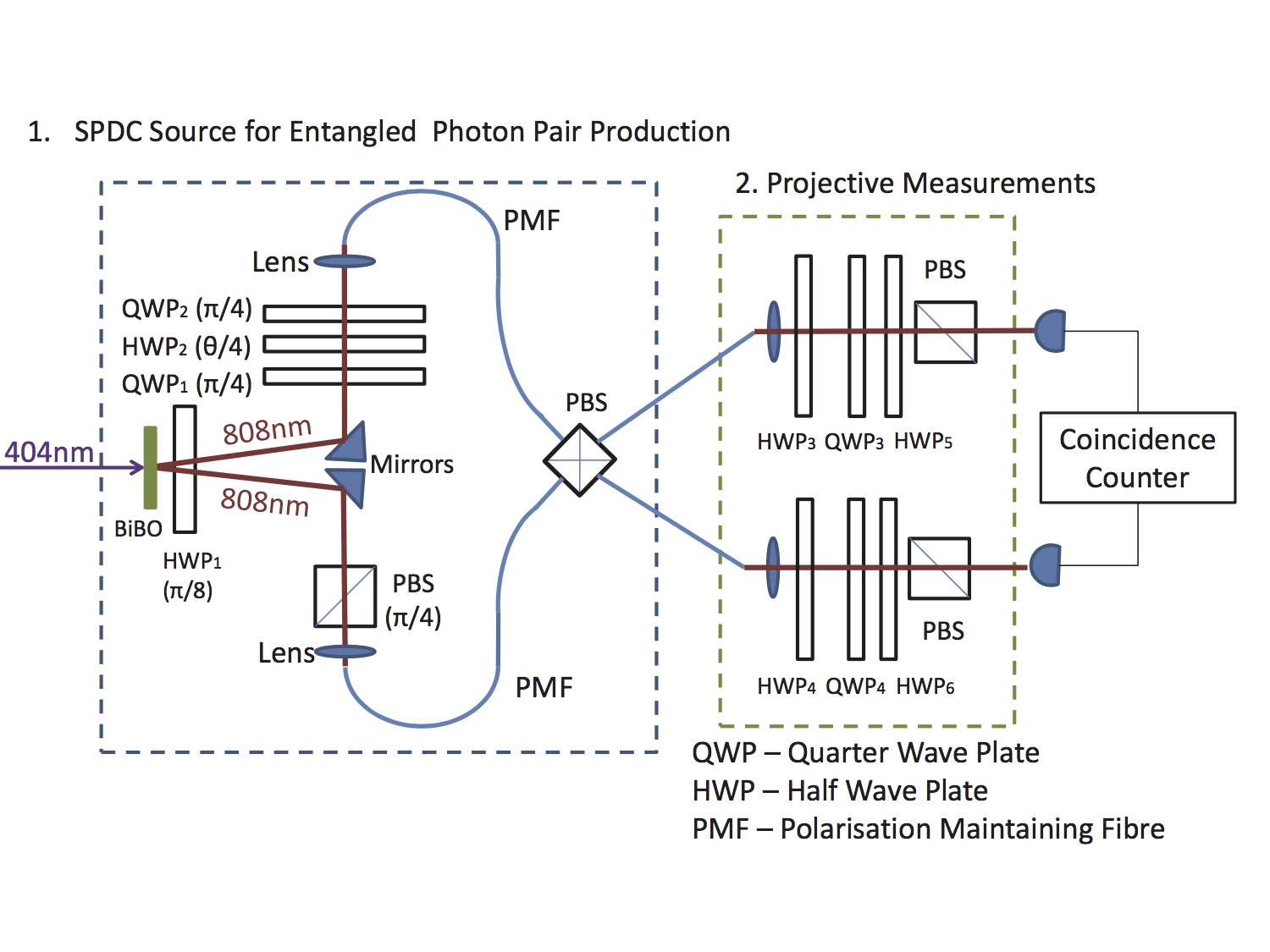}
\caption{\label{fig:scheme}
Scheme for polarisation entangled source. 
Type-I SPDC source is used to generate pairs of H polarised single photons. 
HWP$_1$ rotates $\ket{\text{HH}}$ to $\ket{+}\ket{+}, \ket{+} = \frac{1}{\sqrt{2}}\ket{\text{H}}\ket{\text{V}}$. 
The top arm experiences a polarisation-dependent phase shift, and the state of the bottom arm is purified and matched to the top arm by the PBS rotated by $\pi/4$ radians, which transmits  $\ket{+}$. The photons are then collected into polarisation maintaining fibres (PMF), where the PBS transmits H and reflects V; in combination with post-selection, the measured state is $\frac{1}{\sqrt{2}}(\ket{\text{HV}} + e^{i\theta} \ket{\text{VH}})$ after renormalisation. Standard tomography set up projects the state onto different bases for measurements. See text for further details.  }
\end{figure}

In order to generate polarisation-entangled photon pairs, a set up similar to \cite{matthews2013observing} and identical to \cite{chapman2016experimental} is used.  We use the polarisation encoding where $\ket{\mathrm{H}}\equiv \ket{0}$ and $\ket{\mathrm{V}}\equiv \ket{1}$.
The set up is shown in Fig.\ref{fig:scheme}. A type-I nonlinear crystal (BiBO) is pumped with a vertically polarised cw 
laser (404 nm) at 80 mW, generating pairs of H polarised single photons (808 nm). The photons are spectrally filtered using 3 nm top-hat filteres centred on 808 nm. 
The first half way plate HWP$_1$ (optics axis set to $\pi/8$ radians) rotates the output photon pair from $\ket{\mathrm{HH}}$ to $\ket{\mathrm{++}}$, where $\ket{\mathrm{+}} 
= \frac{1}{\sqrt{2}}\ket{\mathrm{H}}+\ket{\mathrm{V}}$.  The polarisation in the top arm is matched and purified to that of the bottom arm using a polarisation beam splitter (PBS) rotated by $\pi/4$ radians, which transmits $\ket{+}$.

The unitary operation realised by the sequence of wave plates QWP$_1$ + 
HWP$_2(\theta)$ + QWP$_2$, as a function $\theta$ of the HWP$_2$ angle, 
applies a phase to photon in V state, effectively that of a phase gate: 
$\begin{pmatrix}
  1 & 0  \\
  0 & e^{i 4\theta}  \\
 \end{pmatrix}.$
 Denoting the the input photon in the top (bottom) arm with subscript 1 (2), the two-photon state collected into the polarisation mainting fibre (PMF) is therefore  $\frac{1}{\sqrt{2}}(\ket{\text{H}}_1 + e^{i\theta}\ket{\text{V}}_1)\otimes\frac{1}{\sqrt{2}}(\ket{\text{H}}_2 + \ket{\text{V}}_2)$.

We use a silicon avalanche single photon counter and measure in coincidence 
with a time window of 2.5 ns. At the PBS, H is transmitted and V is reflected.
After the fibre PBS crystal, the top (bottom) arm fibre contains the components $(\ket{\text{V}}_1 + \ket{\text{H}}_2)$ 
( $\ket{\text{H}}_1 + e^{i \theta}\ket{\text{V}}_2$). 
However, the design of the fibre PBS is such that the output coupler flips the polarisation state in the bottom arm, therefore
when detected in coincidence, the detector will register either
 $ e^{i\theta}\ket{\text{V}}_1 \ket{\text{H}}_2$ or $\ket{\text{H}}_2 \ket{\text{V}}_1$ , and these two components are made
spatially indistinguishable by adjusting the position of the fibre, giving the post-selected 
state $\frac{1}{\sqrt{2}} (\mathrm{\ket{\text{HV}} + e^{-i \theta}\ket{\text{VH}}})$. HWP$_3$ and 
HWP$_4$ apply the appropriate bit flip operations on each single qubit, and 
QWP$_3$+HWP$_5$ and QWP$_4$+HWP$_6$ project the states onto the different 
bases, which are then measured.

The polarisation maintaining fibres has a beat length (H and V delayed with 
respect to each other by $2\pi$) of 24mm; the coherence length of the photon 
pairs is 250 $\mu$m (extracted from a Hong-Ou-Mandel dip of 99$\%$ 
visibility), which means that if uncompensated, 1.0 m of fibre will spatially 
separate the H and V components of a photon out of coherence, turning a Bell 
state $\frac{1}{\sqrt{2}}\ket{\mathrm{HH+VV}}$ into the classically correlated mixture $
(\mathrm{\frac{\ket{HH}\bra{HH} + \ket{VV}\ket{VV}}{2}})$. 
When producing a Bell state, this spatial decoherence was compensated by 
crossing the slow axis of the PMF and tuning the stage position of the fibre 
(and vice versa when the mixed state is required).

\clearpage

\setcounter{figure}{0}
\makeatletter
\renewcommand{\thefigure}{S\@arabic\c@figure}
\makeatother

\setcounter{equation}{0}
\makeatletter
\renewcommand{\theequation}{S\@arabic\c@equation}
\makeatother

\setcounter{table}{0}
\makeatletter
\renewcommand{\thetable}{S\@arabic\c@table}
\makeatother

\setcounter{page}{1}
\pagebreak

\begin{widetext}
\begin{center}

\textbf{\large High-dimensional entanglement certification}

\textbf{\large Supplemental Materials}

\vspace{3mm}
Zixin Huang$^{1}$, Lorenzo Maccone$^2$, Akib Karim$^{1}$, Chiara Macchiavello$^2$, Robert J. Chapman$^{1}$, Alberto Peruzzo$^{1}$

\vspace{3mm}
\textit{$^1$Quantum Photonics Laboratory, School of Electrical and Computer Engineering, RMIT University, Melbourne, Australia and School of Physics, University of Sydney, NSW 2006, Australia}

\textit{$^2$Dip. Fisica and INFN Sez. Pavia, University of Pavia, via Bassi 6, I-27100 Pavia, Italy}

\end{center}
\end{widetext}

\vspace{3mm}
\subsection{The Fourier basis}

The Fourier basis can be written as single-qubit tensor products in the following way:
\begin{eqnarray}
|f_j\>=\frac 1{\sqrt{d}}\bigotimes_{k=0}^{n-1}(|0\>+\omega^{j2^k}|1\>)
\label{fourierq},
\end{eqnarray}
by expressing $k$ in Eq.(4) in binary form. 
 Using equation~(4), the maximally entangled state in equation~(2) is
perfectly correlated (actually, anti-correlated) in this basis:
\begin{eqnarray}
\tfrac 1{\sqrt{d}}\textstyle{\sum_{j}}|j\>|j\>=\tfrac 1{\sqrt{d}}
\textstyle{\sum_{k}}|f_k\>|f_{-k}\>
\label{antic}.
\end{eqnarray}

\subsection{The $\sigma_x$ basis}

The maxiamlly entangled state
$\ket{00}+\ket{11}=\ket{++}+\ket{--}$, the mapping (2) can be
trivially applied to the $\sigma_x$ basis, namely
\begin{eqnarray}
  \sum_{j}|j\>|j\>&=&(|++\>+|--\>
  )^{\otimes
    n}=\sum_{k}|c_k\>|c_k\>,
\label{mappings2}
\end{eqnarray}
showing that the maximally entangled state in equation~(2) is maximally correlated also in the $\sigma_x$ basis. 

\subsection{Analytical expressions for mutual information}

 On the state $\rho_c$, the joint probabilities
for the computational basis is $p(a_o, b_o)=\delta_{a_ob_o}/d$ with
$\delta_{ab}$ the Kronecker delta. So the mutual information for the
computational basis is $I_{AB}=\log_2 d$ as expected: there is perfect
correlation on such basis in both terms of (6). Writing
$\rho_c$ in the $\sigma_x$ basis, we can calculate the joint
probabilities for it as $p( c_o, d_o)=\delta_{ c_o, d_o}p/d+(1-p)/d^2$
(i.e.~there is still maximal correlation on the entangled part of
$\rho_c$, while there is no correlation on the rest). Whence we can
calculate $I_{CD}$ and find
\begin{eqnarray}
&&I_{AB}+I_{CD}=\log_2d+
\label{i}\\&&\nonumber
\tfrac {(1-p)(d-1)}d\log_2(1-p)+\tfrac
{1+(d-1)p}d\log_2(1+(d-1)p).
\end{eqnarray}
(The same result is obtained also considering the Fourier basis for
$CD$ instead of the $\sigma_x$ basis.) 

Consider now the Werner state $\rho_w$.
 The joint probabilities for
the two complementary observables (computational and Fourier bases)
are respectively
\begin{eqnarray}
  &&p( a_o, b_o)=p/d\delta_{ a_o b_o}+(1-p)/d^2\;,\label{as1}
  \\&& p( c_o,d_o)=p/d\delta_{ a_o,- b_o}+(1-p)/d^2
  \\&&\Rightarrow I_{AB}+I_{CD}=2I_{AB}=\label{wp}\\\nonumber
  &&2\left[\tfrac{1+(d-1)p}d\log_2[1+(d-1)p]+\tfrac{(1-p)(d-1)}d\log_2(1-p)
  \right]  ,
\end{eqnarray}
which is experimentally determined as the green triangles in
Fig.2. 
 The mappings \eqref{antic} and \eqref{mappings2},
and the fact that $\openone$ remains unchanged in any basis implies
that
 $I_{AB}=I_{CD}$ is the same for all the observables considered
here (computational, Fourier and $\sigma_x$). Given the symmetry of
$\rho_w$, our method is able to certify entanglement only for
highly-entangled Werner states.

\subsection{Density matrices}

\begin{figure}[hbt]
\includegraphics[trim = 5cm 0cm 3cm 0cm, clip, width=1.0\linewidth]{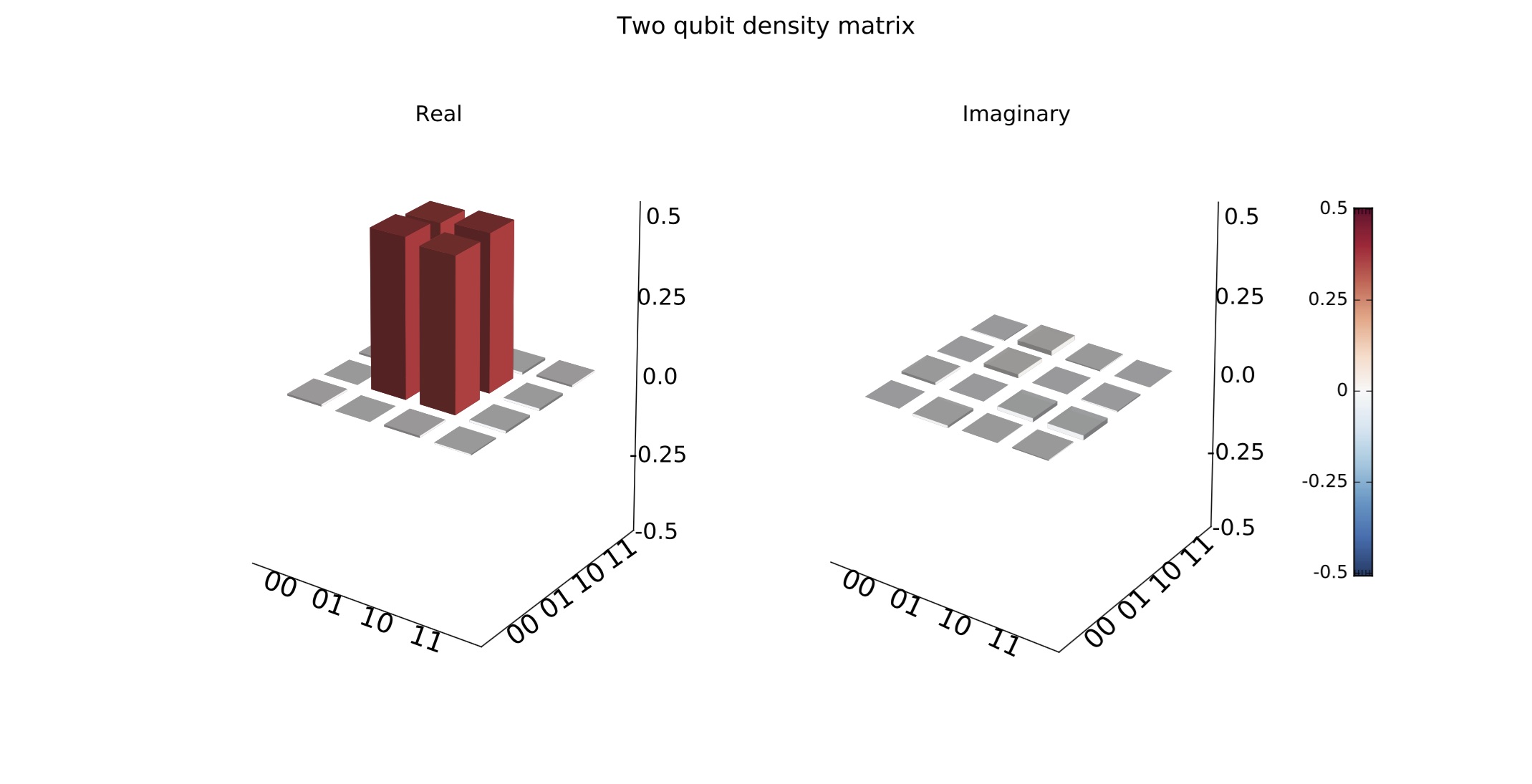}
\caption{\label{f:density_bell_plus} Density matrix for $\frac{1}{\sqrt{2}}(\ket{01}+\ket{10}|)$, fidelity = $97.6 \%$ after maximum likelihood [33] }
\end{figure}

\begin{figure}[hbt]
\includegraphics[trim = 5cm 0cm 3cm 0cm, clip, width=1.0\linewidth]{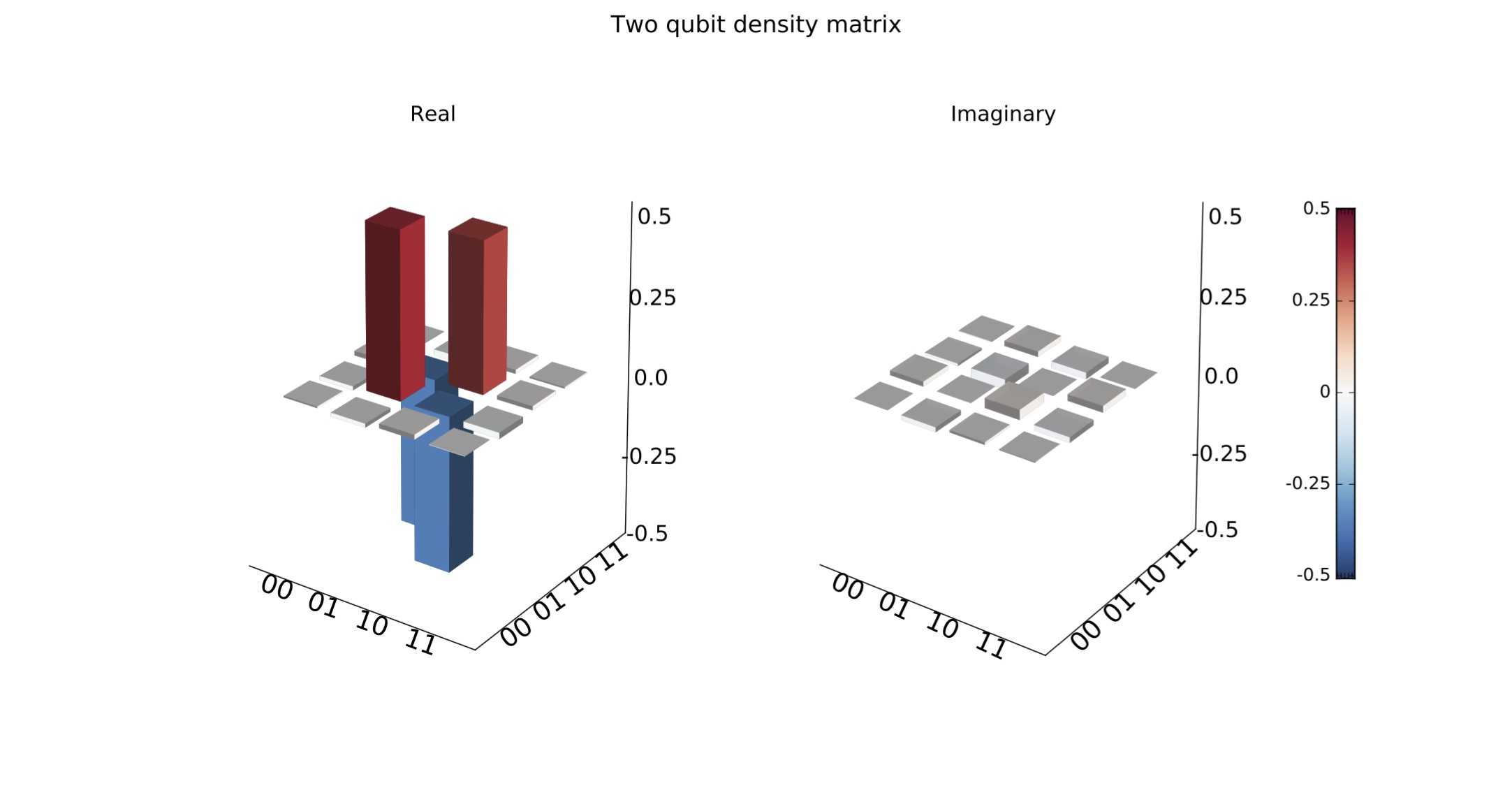}
\caption{\label{f:density_bell_minus} Density matrix for $\frac{1}{\sqrt{2}}(\ket{01}-\ket{10}|)$, fidelity = $97.8 \%$ after maximum likelihood.  }
\end{figure}

\begin{figure}[hbt]
\includegraphics[trim = 5cm 0cm 3cm 0cm, clip, width=1.0\linewidth]{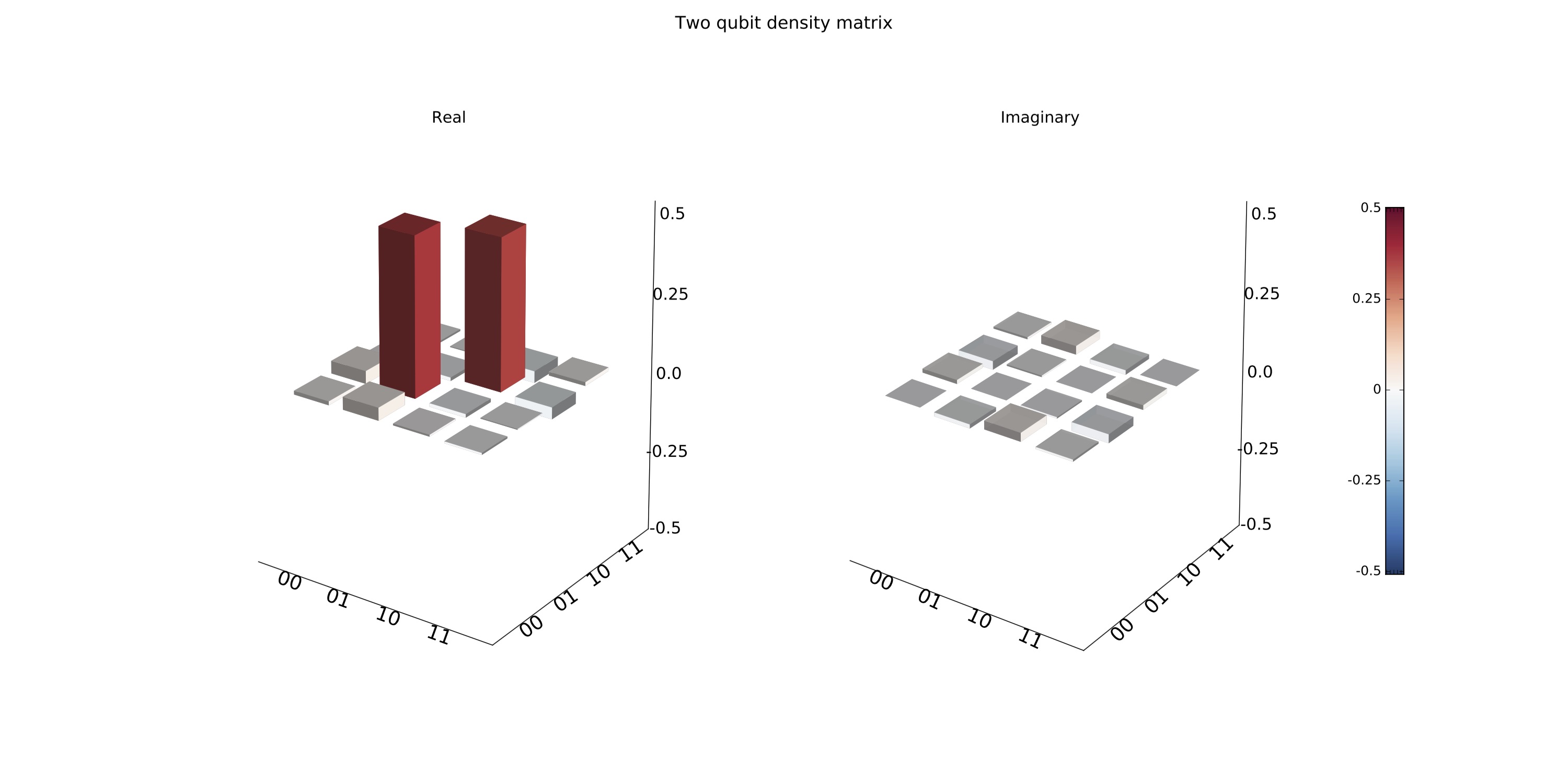}
\caption{\label{f:density_sep} Density matrix for $\frac{\ket{01}\bra{01} + \ket{10}\bra{10}}{2}$, fidelity = $97.6 \%$ after maximum likelihood. }
\end{figure}

The states used in constructing the family of states as in Eq. (6), (7) and (8) are primarily the Bell states $\frac{1}{\sqrt{2}}(\ket{01}+\ket{10}|),\frac{1}{\sqrt{2}}(\ket{01}-\ket{10}|) $ and the decohered, classically correlated state $\frac{\ket{01}\bra{01} + \ket{10}\bra{10}}{2}$. All the others can be accessed by applying the appropriate local unitary transformation.
The density matrices obtained by performing quantum state tomography are shown in Fig.\ref{f:density_bell_plus},\ref{f:density_bell_minus} and \ref{f:density_sep}. The fidelities are given in the captions.

\subsection{$d = 2$ states generation}
$\rho_b$ is generated by averaging measurements for $\ket{\Phi^+}\bra{\Phi^+}$ with phases applied to a single qubit.
$\rho_s$ is generated by decohering the Bell-state $\frac{1}{\sqrt{2}} (\ket{00}+ e^{i\phi}\ket{11})$ to $\frac{\ket{00}\bra{00} + \ket{11}\bra{11} }{2}$ (See Methods), followed by averaging with the appropriate partial bit flip ($\sigma_x \otimes \sigma_x$) applied to both qubits.  For each value of $p$, 40 iterations were run, the upper and lower bounds of the bands are given by the mean of the 40 points $\pm$ 2 standard deviations respectively. The typical count rates were in the order of 800 coincidences/sec, with accidentals subtracted; for each projective measurement, a 10-second integration time was used such that error due to Poissonian noise would account for $\leq 1 \%$ of the measured probability.

$\rho_w$ is made by measuring $\ket{\Phi^+}\bra{\Phi^+}$ and $(\mathbb{I}/4) $ in a time-sharing fashion. All of the states in $d > 2$ are generated by combining pure states measurements.

\section{Data processing and error analysis}

Measurement in each MUB involves projecting the state onto a given basis. For each basis projection onto $\rho_i$, the counts for that projection is normalised with respect to the  normalisation factor ($\mathrm{N_{tot}}$), given :
$\mathrm{N_{tot}}= \bra{00}\rho_i \ket{00} + \bra{01}\rho_i \ket{01} + \bra{10}\rho_i \ket{10} + \bra{11}\rho_i \ket{11}$ 

Each statistic was assumed to be a Poissonian process, with standard deviation $\Delta \mathrm{N_i}$ = $\sqrt{\mathrm{N_i}}$. The error bars displays 2 stdev with the following method of analysis.

$\mathrm{p_i = N_i/N_{tot}}$, therefore $\mathrm{\Delta p_i = p_i (\frac{\Delta N_i}{N_{tot}} + \frac{\Delta N_{tot}}{N_{tot}}})$
eg. with a total of 10000 counts, the following shows the basis vector, counts for the projection and probability:
\begin{equation}
  \begin{pmatrix}
       HH \\
       HV \\
       VH \\
       VV 
   \end{pmatrix} \rightarrow
  \begin{pmatrix}
     1000 \\
       4000 \\
       4500 \\
        500 
   \end{pmatrix} \rightarrow
   \begin{pmatrix}
     0.10 \\
       0.40 \\
       0.45 \\
       0.05 
   \end{pmatrix}
\end{equation}

\noindent
The Shannon entropy for qubit one is given by:
\begin{equation}
\mathrm{H(A) = - p(H)log2(p(H)) + p(V)log2p(V)}
\end{equation}
\noindent ie
\begin{eqnarray}
\mathrm{H(A)} &=& - ( (0.1+0.4) \mathrm{log2} (0.1+0.4)   \nonumber\\
              & & + (0.45+0.5)  \mathrm{log2}(0.45+0.45) 
\end{eqnarray}

\begin{equation}
  \begin{split}
    \mathrm{ H(A|B) = - ( (0.1) log2 (\frac{0.1}{0.1+0.45} ) + (0.45) log2( \frac{0.45}{0.1+0.45})} \\
    + \mathrm {(0.40) log2 (\frac{0.40}{0.40+0.05}) + (0.05) log2 (\frac{0.40}{0.40+0.05}) }
  \end{split}
\end{equation}

The error in the probabilities are therefore:

\begin{equation}
\Delta p_i = \begin{pmatrix}
         0.1 (\frac{\sqrt{1000}}{1000} + \frac{\sqrt{10000}}{10000}) \\
        0.40 (\frac{\sqrt{4000}}{4000} + \frac{\sqrt{10000}}{10000})  \\
        0.45 (\frac{\sqrt{4500}}{4500} + \frac{\sqrt{10000}}{10000})  \\
        0.05 (\frac{\sqrt{500}}{500} + \frac{\sqrt{10000}}{10000})  
       \end{pmatrix}
\end{equation}

\begin{equation}
\mathrm{I} = - \sum_{i} \mathrm{p_i log_2 (p_i) }
\end{equation}

At each point displayed on the graphs in the main text, the error bars were calculated via error propagation. The uncertainty in information, $\Delta I$ then follows as:

\begin{equation}
\mathrm{\Delta I} = - \mathrm{\frac{1}{ln(2)} \sum_{i}  ( \Delta p_i ln(p_i) +  p_i  \Delta(ln(p_i) ) )} \
\end{equation}

\begin{equation}
=- \mathrm{ \frac{1}{ln(2)} \sum_{i}  ( \Delta p_i ln(p_i) + 1 ) )}
\end{equation}

\section{Maximal MUB's in $d=3$ and 4}

We used the MUB's from [34] to perform our calculations.  Note that for d = 4, out of the four sets of MUB's provided, there is an error in that the two sets are identical. We amended this, and the ones we used are specified in Eq.\eqref{eq:fourMUBs}

For $d=3$:

\begin{eqnarray}
\{\ket{F3_a} \} &=& \{ \frac{1}{3}[ 1,1,1    ] ,\frac{1}{3}[ 1, \omega, \omega^2    ] , \frac{1}{3}[ 1, \omega^2, \omega    ]     \}  \nonumber\\
\{\ket{F3_b} \} &=& \{ \frac{1}{3}[ 1, \omega, \omega     ] ,\frac{1}{3}[  1, \omega^2, 1   ] , \frac{1}{3}[ 1,1, \omega^2    ]     \} \nonumber\\
\{\ket{F3_c} \} &=& \{ \frac{1}{3}[ 1, \omega^2, \omega^2    ] ,\frac{1}{3}[ 1, \omega, 1    ] , \frac{1}{3}[ 1,1,\omega    ]       \}
\end{eqnarray}

\noindent where $\omega= e^{2\pi i/d}$. Together with the computational basis, these form the four sets of MUB's in $d=3$.

For the computational and the Fourier basis ($\{\ket{F3_a}\}$), maximal correlation occur when the two systems are measured in these the same respective bases. For $\{\ket{F3_b} \}$ and $\{\ket{F3_c} \}$ however, there is zero correlation if both sides are measured in the same bases. To achieve maximum correlation, when system one is measured in $\{\ket{F3_b} \}$ ( $\{\ket{F3_c} \}$), system two must be measured with a different set of MUB as defined in Eqn.\eqref{eq:f3bc} (\eqref{eq:f3cc}).

\begin{figure}[hbt]
\includegraphics[trim = 0cm 0cm 0cm 0cm, clip, width=0.9\linewidth]{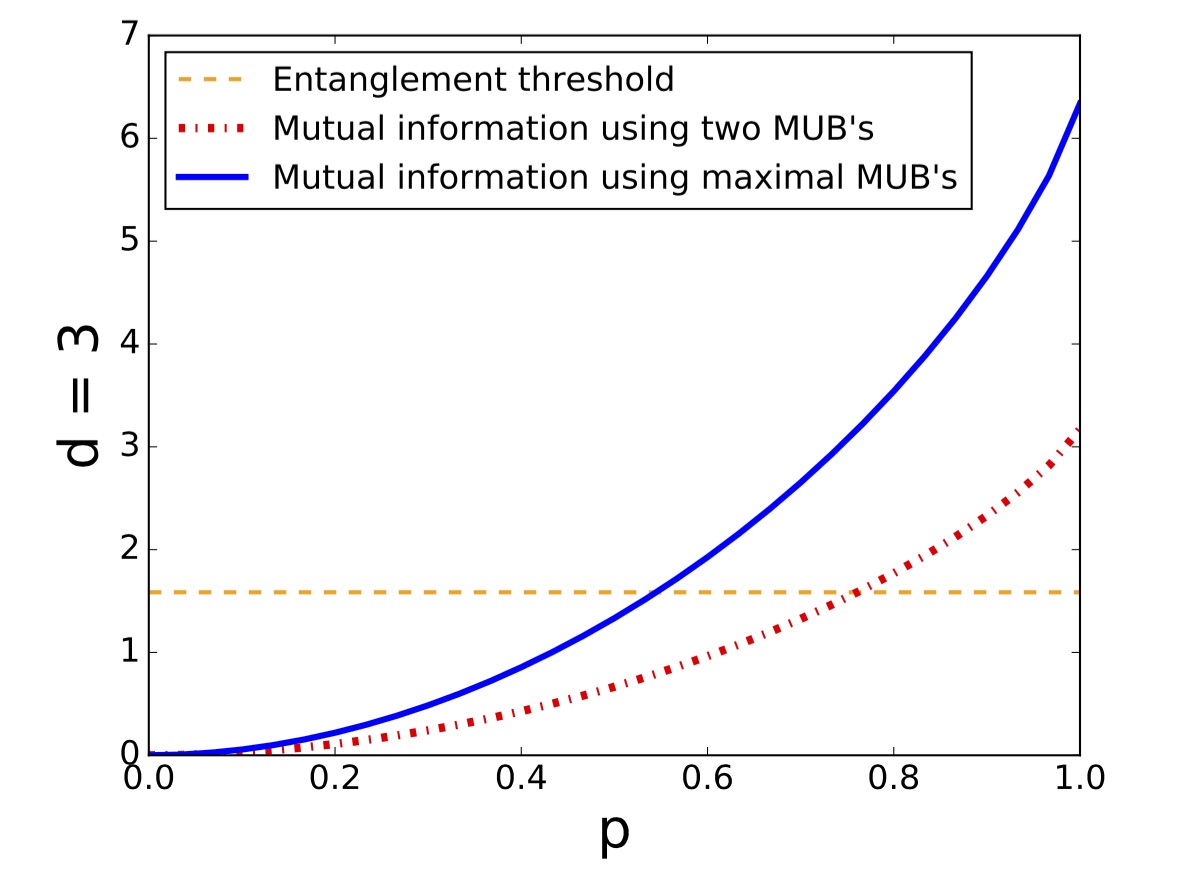}
\caption{\label{f:qutritmaximal} Mutual information of the Werner state in Eqn (7) for $d=3$ when measured in only two MUB's (red dotted line), or all five sets of MUB's (blue solid line)}
\end{figure}

Comparing the mutual information for using just two MUB's (red dotted) and the full four MUB's, are given in Fig.\ref{f:qutritmaximal}.

\begin{eqnarray}
\{ \frac{1}{3}[ \omega,1,1]  , \frac{1}{3}[1,1,\omega]    , \frac{1}{3}[1,\omega,1] \}
\label{eq:f3bc}
\end{eqnarray}

\begin{eqnarray}
\{ \frac{1}{3}[ \omega^2,1,1]  , \frac{1}{3}[1,1,\omega^2]    , \frac{1}{3}[1,\omega^2,1] \}
\label{eq:f3cc}
\end{eqnarray}

\begin{figure}[hbt]
\includegraphics[trim = 0cm 0cm 0cm 0cm, clip, width=0.9\linewidth]{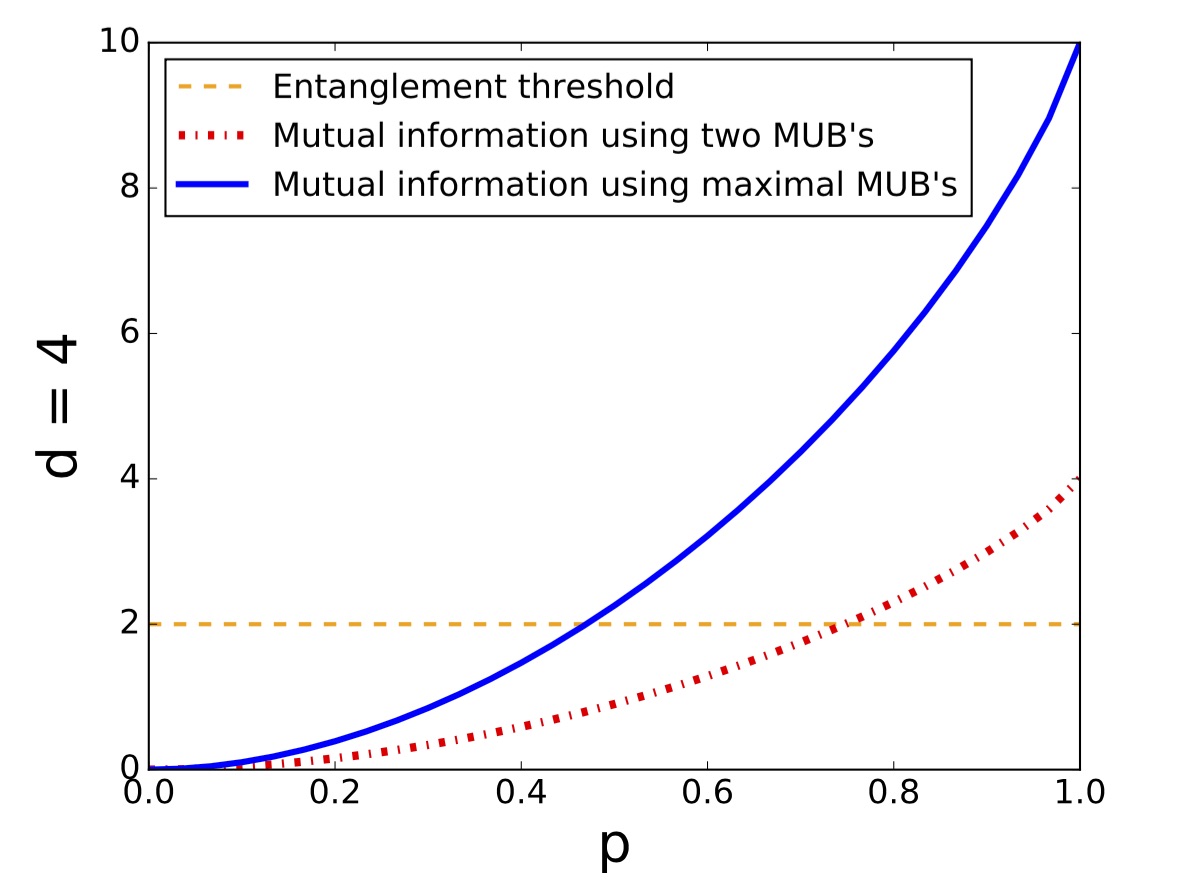}
\caption{\label{f:ququartmaximal} Mutual information of the Werner state in Eqn (7) for $d=4$ when measured in only two MUB's (red dotted line), or all five sets of MUB's (blue solid line)}
\end{figure}

For $d=4$, we used the following bases:
\begin{eqnarray}
\{\ket{F4_a} \} &= &\{ \frac{1}{2}[ 1,1,1,1,    ],  \frac{1}{2}[ 1,1,-1,-1 ] ,   \nonumber\\ \nonumber \label{eq:fourMUBs}
                &  &   \frac{1}{2}[ 1,-1,-1,1   ],  \frac{1}{2}[ 1,-1,1,-1 ]   \} \\ \nonumber
\{\ket{F4_b} \} &=& \{ \frac{1}{2}[ 1,-1,-i,-i  ] , \frac{1}{2}[ 1,-1,i,i    ] , \\ \nonumber
                & &    \frac{1}{2}[ 1,1,i,-i    ] , \frac{1}{2}[ 1,1,-i,i    ]   \} \\ \nonumber
\{\ket{F4_c} \} &=& \{ \frac{1}{2}[ 1,-i,-i,-1  ] , \frac{1}{2}[ 1,-i,i,1    ] , \\ \nonumber
                & &    \frac{1}{2}[ 1,i,i,-1    ] , \frac{1}{2}[ 1,i,-i,1    ]   \} \\ \nonumber
\{\ket{F4_d} \} &=& \{ \frac{1}{2}[ 1,-i,1,-i   ] , \frac{1}{2}[ 1,-i,1,i    ] , \\ 
                & &    \frac{1}{2}[ 1,i,1,-i    ] , \frac{1}{2}[ 1,i,-1,i    ]    \} 
\end{eqnarray}
Together with the computational basis, these form the five sets of MUB's in $d=4$.  The mutual information for the Werner state, when measured using two MUB's (red dotted) and the full five MUB's, are given in Fig.\ref{f:ququartmaximal}.

\end{document}